\documentclass[12pt]{article}

\usepackage{amsmath}

\usepackage[dvipdfmx]{graphicx}

\usepackage{subfig}

\usepackage{array}

\usepackage{amsfonts}

\usepackage{epsf}

\usepackage{epsfig}

\usepackage{amssymb}

\usepackage{bbm}

\usepackage{delarray}

\usepackage{caption}

\usepackage{amstext}

\usepackage{graphics}

\usepackage{epstopdf}

\usepackage[titletoc,toc,title]{appendix}

\catcode`ð=\active
 \defð{\u{g}}
 \catcode`Ð=\active
 \defÐ{\u{G}}
 \catcode`Ý=\active
 \defÝ{\. I}
 \catcode`ö=\active
 \defö{\"{o}}
 \catcode`Ö=\active
 \defÖ{\"O}
 \catcode`ü=\active
 \defü{\"{u}}
 \catcode`Ü=\active
 \defÜ{\"{U}}
 \catcode`Þ=\active
 \defÞ{\c{S}}
 \catcode`þ=\active
 \defþ{\c{s}}
 \catcode`ý=\active
 \defý{{\i}}
 \catcode`ç=\active
 \defç{\d{c}}
 \catcode`Ç=\active
 \defÇ{\d{C}}

\numberwithin{equation}{section}
\begin{document}
\title{\textbf{Solution of Effective-Mass Dirac Equation with Scalar-Vector and Pseudoscalar Terms for Generalized Hulth\'{e}n Potential}}

\author{{\small Altu\u{g} Arda$^a$\footnote{Present adress: Department of Mathematical Science, City University London, UK}$^c$\footnote{arda@hacettepe.edu.tr}}\\
$^a${\small \emph{Department of Physics Education, Hacettepe University}}, \\
{\small \emph{06800,
Ankara, Turkey}}\\
 \\}
\date{}

\maketitle

\begin{abstract}
We find the exact bound-state solutions and normalization constant for the Dirac equation with scalar-vector-pseudoscalar interaction terms for the generalized Hulth\'{e}n potential in the case where we have a particular mass function $m(x)$. We also search the solutions for the constant mass where the obtained results correspond to the ones when the Dirac equation has spin and pseudospin symmetry, respectively. After giving the obtained results for the non-relativistic case, we search then the energy spectra and corresponding upper and lower components of Dirac spinor for the case of $PT$-symmetric forms of the present potential.

PACS: 03.65.-w, 03.65.Pm, 03.65.Ge

Keywords: Dirac equation, scalar-vector-pseudoscalar term, Hulth\'{e}n potential, $PT$-symmetry, Nikiforov-Uvarov method
\end{abstract}
\newpage

\section{Introduction}

The Hulth\'{e}n potential [1] is one of the best known potentials in physics, as a short-range potential [2]. In the present work, we deal with the following form [3, 4]
\begin{eqnarray}
V(x)=V_{0}\,\frac{e^{-2\beta x}}{qe^{-2\beta x}-1}\,,
\end{eqnarray}
where the parameter $V_{0}$ can be written as $Z(2\beta)$ with the constant $Z$, and $\beta$ is the screening parameter (in atomic units) [2] with deformation parameter $q$. The constant $Z$ is related with the atomic number if one uses this potential in atomic physics. Basically, the  Hulth\'{e}n potential is a special form of the Eckart potential [2, 5].

As a short-range potential, the Hulth\'{e}n potential has a great advantage because the Schr\"{o}dinger equation can be solved exactly for this potential with $\ell=0$. With this advantage, the Hulth\'{e}n potential has been used in different areas of physics, such as in solid-state physics [6], nuclear and particle physics [7], atomic physics [8], and chemical physics [9], and investigated with various techniques [2, 10-17].

In this work, we search the bound state solutions of the generalized Hulth\'{e}n potential which can be written also in a complex form identifying the $PT$-symmetric case in a closed form for the case where the mass depends on spatially coordinate, and extend the Dirac equation including the scalar, vector and pseudoscalar interaction terms to this case. After the works by von Roos [18], and Levy-Leblond [19], the solutions of relativistic and non-relativistic wave equations with a position-dependent mass have received great attention in literature [3, references therein]. In Ref. [20], the bound state solutions of the Klein-Gordon (KG) and Dirac equations with the Hulth\'{e}n potential by using the approach proposed by Biedenharn have been worked where the scattering state solutions have been also presented. In Ref. [21], the analytical results for the bound states of the Dirac equation with the generalized Hulth\'{e}n potential as a tensor term have been studied within the concept of the SUSYQM. In the present work, we extend the search including the solutions of the Dirac equation having a pseudoscalar interaction term, as in Refs. [33-36], for the $q$-parameter Hulth\'{e}n potential within the position-dependent mass (PDM) formalism. This formalism gives an opportunity such as writing the analytical results for the case where the mass is constant. This means that our results are also available for the cases where the Dirac equation has pseudospin and spin symmetry. Our generic results will be given in below makes it possible to give the "wave functions" with their normalization constants both for the cases of PDM formalism and constant mass. We search here also the analytical results for the $PT$-symmetric/non-Hermitian and $PT$-symmetric/pseudo-Hermitian form of the Hulth\'{e}n potential for both of upper and lower component which are presented again within the PDM formalism. These give us also the results for the case where the mass is constant if necessary. Among the above results, because of the $q$-parameter in potential, we apply our results for three different form of the potential as special cases.

The organization of this work is as follows. In Section 2, we write the Dirac equation with scalar ($V_{S}(x)$), vector ($V_{V}(x)$), and pseudoscalar ($V_{P}(x)$) potentials in $1+1$ dimension for the case where the mass is a function of spatially coordinate. In Section 3, we search the bound-state solutions for upper and lower component of the Dirac spinor separately, and give the normalization constant. We construct a relation between the mass function and the potentials to reduce the Dirac equation to an analytically solvable form of second order differential equation. We give also the results for the case where the mass is constant, and observe that the results obtained for this case correspond to the solutions when the spin and pseudospin symmetry occur in Dirac equation. The spin symmetry appears when the difference of the scalar and vector potentials is constant, i.e., $\Delta(x)=const.$, and the pseudospin symmetry appears when the sum of the scalar and vector potentials is constant, i.e., $\Sigma(x)=const.$ [20-22]. Finally, we obtain the non-relativistic result for the bound-state solution for the generalized Hulth\'{e}n potential. The present work can also be seen as an application of the parametric generalization of the Nikiforov-Uvarov method which will be given in Appendix briefly [23, 24]. In Section 4, we write the generalized Hulth\'{e}n potential in a complex form which corresponds to the $PT$-symmetric form of the potential, and find the energy levels for upper and lower component of the Dirac spinor with normalized wave functions. The $PT$-symmetric formulation with non-Hermitian Hamiltonians having real or complex spectra of quantum mechanics has received a great attention in literature after the work by Bender and Boettcher [25-27]. In Section 5, we collect briefly our analytical results for special values of the parameter $q$ corresponding to the standard Hulth\'{e}n potential ($q=1$), to the Woods-Saxon potential ($q=-1$), and to the exponential potential ($q=0$) while the mass depends on spatially coordinate. We give our conclusions in last Section.

\section{Dirac Equation in $1+1$ Dimension}

The time-independent Dirac equation for a spin-$1/2$ particle subjected to scalar, vector and pseudoscalar potentials in terms of the $\Sigma(x)=V_{V}(x)+V_{S}(x)$, and $\Delta(x)=V_{V}(x)-V_{S}(x)$ is given by ($\hbar=c=1$) [28-34]
\begin{eqnarray}
\left(\sigma_{1}p+\sigma_{3}m(x)+\frac{1+\sigma_{3}}{2}\,\Sigma(x)+\frac{1-\sigma_{3}}{2}\,\Delta(x)+\sigma_{2}V_{P}(x)\right)\Psi(x)=E\Psi(x)\,,
\end{eqnarray}
where $\sigma_{1}, \sigma_{2}$ and $\sigma_{3}$ are the Pauli spin matrices, and we write the mass as $m(x)$. By taking the Dirac spinor as $\Psi=(\phi_{1}, \phi_{2})^{t}$ where $t$ indicates the transpose, we obtain the following first order coupled equations for upper and lower components
\begin{eqnarray}
&&-i\frac{d\phi_{1}(x)}{dx}-[m(x)-\Delta(x)+E]\phi_{2}(x)+iV_{P}\phi_{1}(x)=0\,,\\
&&-i\frac{d\phi_{2}(x)}{dx}+[m(x)+\Sigma(x)-E]\phi_{1}(x)-iV_{P}\phi_{2}(x)=0\,,
\end{eqnarray}

Writing $\phi_{2}(x)$ in terms of $\phi_{1}(x)$ with the help of Eq. (2.2), and inserting it into Eq. (2.3) gives us
\begin{eqnarray}
&&\left(\frac{d^{2}\phi_{1}(x)}{dx^{2}}-\frac{dV_{P}(x)}{dx}\phi_{1}(x)-V_{P}(x)\frac{d\phi_{1}(x)}{dx}\right)[m(x)-\Delta(x)+E]\nonumber\\&-&\left(\frac{dm(x)}{dx}-\frac{d\Delta(x)}{dx}\right)\left(\frac{d\phi_{1}(x)}{dx}-V_{P}(x)\phi_{1}(x)\right)\nonumber\\&+&V_{P}(x)\left(\frac{d\phi_{1}(x)}{dx}-V_{P}(x)\phi_{1}(x)\right)[m(x)-\Delta(x)+E]\nonumber\\&-&[m(x)-\Delta(x)+E]^{2}[m(x)+\Sigma(x)-E]\phi_{1}(x)=0\,,\nonumber
\end{eqnarray}
We write here the equality $dm(x)/dx=d\Delta(x)/dx$ between the mass function and the potentials to reduce the above complicated equation to a simpler one which can be solved analytically. This relation gives also us the opportunity about finding the mass function explicitly, and the second order equation for upper component $\phi_{1}(x)$ as
\begin{eqnarray}
\left\{\frac{d^2}{dx^2}-\frac{dV_{P}(x)}{dx}-V^{2}_{P}(x)-[m(x)+E-\Delta(x)][m(x)+\Sigma(x)-E]\right\}\phi_{1}(x)=0\,,
\end{eqnarray}

By following similar steps, and using the equality for the mass function as $dm(x)/dx=-d\Sigma(x)/dx$, we obtain the second order equation for lower component $\phi_{2}(x)$ as
\begin{eqnarray}
\left\{\frac{d^2}{dx^2}+\frac{dV_{P}(x)}{dx}-V^{2}_{P}(x)-[m(x)+E-\Delta(x)][m(x)+\Sigma(x)-E]\right\}\phi_{2}(x)=0\,.
\end{eqnarray}

Eqs. (2.4) and (2.5) can be solved by using the parametric generalization of the Nikiforov-Uvarov method which is given in Appendix briefly. In the next Section, we solve the above equations for the scalar, vector and pseudoscalar potentials by identifying them in terms  of the Hulth\'{e}n potential given in Eq. (1.1). First, we find the appropriate mass function by using the equalities, and then we write the bound-state solutions with the corresponding normalized wave functions.

\section{Bound States for generalized Hulth\'{e}n Potential}

We are now in a position to identify the potentials in terms of the generalized Hulth\'{e}n potential. We tend to write them as following [3, 4]
\begin{eqnarray}
&&V_{V}(x)=V_{0}\,\frac{e^{-2\beta x}}{qe^{-2\beta x}-1}\,,\nonumber\\
&&V_{S}(x)=-S_{0}\,\frac{e^{-2\beta x}}{qe^{-2\beta x}-1}\,,\nonumber\\
&&V_{P}(x)=V_{i}\,\frac{e^{-2\beta x}}{qe^{-2\beta x}-1}\,,
\end{eqnarray}
where $i=1, 2$, and $V_{1}$ for the upper component $\phi_{1}(x)$, $V_{2}$ for the lower component $\phi_{2}(x)$. We obtain Eq. (2.5) for $\phi_{2}(x)$ by replacement $V_{2} \leftrightarrow -V_{1}$ in Eq. (2.4). In addition, we can handle the explicit form of the wave function for $\phi_{2}(x)$ by doing $\beta \leftrightarrow -\beta$ in $\phi_{1}(x)$.

From Eq. (3.1), we have
\begin{eqnarray}
\Sigma(x)=(V_{0}-S_{0})\,\frac{e^{-2\beta x}}{qe^{-2\beta x}-1}\,\,;\,\,\,\Delta(x)=(V_{0}+S_{0})\,\frac{e^{-2\beta x}}{qe^{-2\beta x}-1}\,,
\end{eqnarray}
By using Eqs. (3.1) and (3.2), we write the mass function from the equality obtained for $\phi_{1}(x)$ as $m(x)=m_{0}+m_{1}\,\frac{e^{-2\beta x}}{qe^{-2\beta x}-1}$, where the parameter $m_{0}$ is basically the integral constant, and we denote it as 'constant mass', the other parameter $m_{1}$ is obtained as $m_{1}=V_{0}+S_{0}$. This means that the mass parameter $m_{1}$ contains the contributions coming from vector and scalar potentials. The equality obtained for $\phi_{2}(x)$ gives us the mass function as $m(x)=m_{0}+m_{2}\,\frac{e^{-2\beta x}}{qe^{-2\beta x}-1}$ with $m_{2}=V_{0}-S_{0}$ including the contributions coming from vector and scalar potentials. So, we can combine these two mass functions in a single form as $m(x)=m_{0}+m_{i}\,\frac{e^{-2\beta x}}{qe^{-2\beta x}-1}$ which will be used in computation below.

By using a new variable as $s=1/(1-qe^{-2\beta x})$ ($-\infty<x<+\infty \rightarrow 0\leqslant s \leqslant 1$), using Eqs. (3.1)-(3.2), and inserting the mass function $m(x)$, we have the following representative equation for both components
\begin{eqnarray}
\left\{\frac{d^2}{ds^2}+\frac{1-2s}{s(1-s)}\frac{d}{ds}-\frac{1}{s^2(1-s)^2}\left[\frac{A}{4\beta^2}+\frac{2B}{4\beta^2}s+\frac{C}{4\beta^2}s^2\right]\right\}\phi_{i}(s)=0\,,
\end{eqnarray}
where
\begin{eqnarray}
&&A=m^{2}_{0}-E^{2}+2QV_{0}(E+m_{0})+a_{i}\,,\nonumber\\
&&B=-QV_{0}(E+m_{0})-b_{i}\,,\nonumber\\
&&C=2\beta QV_{i}+Q^{2}V^{2}_{i}+Q^{2}[(m_{i}-S_{0})^{2}-V^{2}_{0}]\,,
\end{eqnarray}
with
\begin{eqnarray}
&&a_{i}=2Qm_{0}m_{i}+Q^{2}V^{2}_{i}+Q^{2}[(m_{i}-S_{0})^{2}-V^{2}_{0}]-2Qm_{0}(V_{0}+S_{0})\,,\nonumber\\
&&b_{i}=QV_{i}(\beta+QV_{i})+Q^{2}[(m_{i}-S_{0})^{2}-V^{2}_{0}]-Qm_{0}(S_{0}+V_{0})+Qm_{0}m_{i}\,.
\end{eqnarray}
and $Q=1/q$. Eq. (3.3) can be solved by using the parametric Nikiforov-Uvarov method. For this aim, we compare Eq. (3.3) with Eq. (A.1) in Appendix, and with the help of Eq. (A.3) we obtain the parameter set
\begin{eqnarray}
\alpha_{1}=1, \alpha_{2}=2,\alpha_{3}=1,\xi_{1}=\frac{C}{4\beta^2},\xi_{2}=-\frac{2B}{4\beta^2},\xi_{3}=\frac{A}{4\beta^2},\nonumber\\\alpha_{4}=\alpha_{5}=0,\alpha_{6}=\xi_{1},\alpha_{7}=-\xi_{2},\alpha_{8}=\xi_{3},\alpha_{9}=\xi_{1}-\xi_{2}+\xi_{3},\nonumber\\\alpha_{10}=1+2\sqrt{\xi_{3}\,},\alpha_{11}=2+2(\sqrt{\xi_{1}-\xi_{2}+\xi_{3}\,}+\sqrt{\xi_{3}\,}),\nonumber\\
\alpha_{12}=\sqrt{\xi_{3}\,},\alpha_{13}=-(\sqrt{\xi_{1}-\xi_{2}+\xi_{3}\,}+\sqrt{\xi_{3}\,})\,,
\end{eqnarray}

With the help of Eq. (A.2) in Appendix, we write the energy spectrum of the Dirac equation with scalar-vector-pseudosclar generalized Hulth\'{e}n potential within the position-dependent mass formalism as
\begin{eqnarray}
&\left[\sqrt{m^{2}_{0}-E^{2}+2QV_{0}(E+m_{0})+a_{i}\,}+\sqrt{m^{2}_{0}-E^{2}\,}+\beta(2n+1)\right]^{2}\nonumber\\&-\left[(\beta+QV_{i})^{2}+Q^{2}[(m_{i}-S_{0})^{2}-V^{2}_{0}]\right]=0\,,
\end{eqnarray}
which can be solved numerically to get the energy eigenvalues.

In order to handle the generic wave function for the upper and lower component of the Dirac spinor, we use Eq. (A.4) in Appendix which gives
\begin{eqnarray}
\phi_{i}(s)=N_{i}s^{\alpha'/2}(1-s)^{\beta'/2}P_{n}^{(\alpha',\beta')}(1-2s)\,.
\end{eqnarray}
with $\alpha'=2\sqrt{\xi_{3}\,}$, $\beta'=2\sqrt{\xi_{1}-\xi_{2}+\xi_{3}\,}$, and the normalization constant $N_{i}$. Let us now find the normalization constant. Using a new variable $z=1-2s$ ($0\leqslant s \leqslant 1 \rightarrow +1\leqslant z\leqslant -1$), and writing the normalization condition $\int_{-\infty}^{+\infty}|\phi_{i}(x)|^{2}dx=1$ as
\begin{eqnarray}
\int_{-1}^{+1}|\phi_{i}(z)|^{2}\frac{dz}{\beta(1-z)(1+z)}=1\,,
\end{eqnarray}
we get
\begin{eqnarray}
\frac{|N_{i}|^{2}}{\beta}\frac{1}{2^{\alpha'+\beta'}}\int_{-1}^{+1}(1-z)^{\alpha'-1}(1+z)^{\beta'-1}\left[P_{n}^{(\alpha',\beta')}(z)\right]^{2}dz=1\,,
\end{eqnarray}
By using the following representation of the Jacobi polynomials [33]
\begin{eqnarray}
P_{n}^{(\alpha',\beta')}(z)=\frac{1}{n!}\sum_{\ell=0}^{n}\,\frac{1}{\ell!}\,(-n)_{\ell}(n+\alpha'+\beta'+1)_{\ell}(n+\alpha'+1)_{\ell}\left(\frac{1-z}{2}\right)^{\ell}\,,
\end{eqnarray}
Eq. (3.10) is written as
\begin{eqnarray}
\frac{|N_{i}|^{2}}{\beta}\frac{1}{2^{\alpha'+\beta'+\ell}}\frac{1}{n!}\sum_{\ell=0}^{n}\,\frac{1}{\ell!}\,(-n)_{\ell}(n+\alpha'+\beta'+1)_{\ell}(n+\alpha'+1)_{\ell}\nonumber\\ \times\int_{-1}^{+1}(1-z)^{\alpha'-1+\ell}(1+z)^{\beta'-1}P_{n}^{(\alpha',\beta')}(z)dz=1\,,
\end{eqnarray}
where $(n)_{r}$ is the Pochammer symbol [35]. With the help of the following integral equation including a Jacobi polynomial written in terms of the hypergeometric function $\,_{3}F_{2}(-n, a, b; c,d;y)$ [35]
\begin{eqnarray}
\int_{-1}^{+1}(1-y)^{\rho}(1+y)^{\sigma}P_{n}^{(\alpha',\beta')}(y)dy=\frac{2^{\rho+\sigma+1}\Gamma(\rho+1)\Gamma(\sigma+1)\Gamma(n+1+\alpha')}{n!\Gamma(\rho+\sigma+2)\Gamma(\alpha'+1)}\nonumber\\ \times
\,_{3}F_{2}(-n, n+\alpha'+\beta'+1,\rho+1;\alpha'+1, \rho+\sigma+2;1)\,,
\end{eqnarray}
with the conditions $Re\rho > -1$, and $Re \sigma > -1$, the normalization constant is computed
\begin{eqnarray}
N_{i}=\sqrt{\Gamma'\Gamma''\,}\,,
\end{eqnarray}
where
\begin{eqnarray}
\Gamma'&=&\frac{2\beta(n!)^{2}\Gamma(\alpha'+\beta')\Gamma(\alpha'+1)}{\Gamma(\beta)\Gamma(n+\alpha'+1)}\,,\nonumber\\
\Gamma''&=&\left[\sum_{\ell=0}^{n}\,\frac{1}{\ell!}\,(-n)_{\ell}(n+\alpha'+\beta'+1)_{\ell}(n+\alpha'+1)_{\ell}\right.\nonumber \\ &&\left.\,_{3}F_{2}(-n, n+\alpha'+\beta'+1,\alpha'+\ell;\alpha'+1, \alpha'+\beta'+\ell;1)\right]^{-1}\,.\nonumber
\end{eqnarray}
The condition to be satisfied in Eq. (3.13) gives an upper limit for $\ell$ as $\ell < \frac{A}{\beta}$ which can be used to determining the greatest integer value for the quantum number $n$ as $n < \left[\frac{A}{\beta}\right]$.

We obtain the formal analytical solutions for the problem under consideration giving the results in terms of representative equations (3.7) and (3.8). Now we move on to consider the upper and lower components of the Dirac spinor separately, and summarize the results for the case where the mass is constant, and the case of non-relativistic limit for the present problem.

\subsection{Results}

For the upper component ($i=1$), we write the potential parameter as $m_{1}=V_{0}+S_{0}$ which gives the following energy eigenvalue equation from (3.7) as
\begin{eqnarray}
\left[\sqrt{m^{2}_{0}-E^{2}+2QV_{0}(E+m_{0})+a_{1}\,}+\sqrt{m^{2}_{0}-E^{2}\,}+\beta(2n+1)\right]^{2}-(\beta+QV_{1})^{2}=0\,,
\end{eqnarray}
with $a_{1}=Q^{2}V^{2}_{1}$, and $b_{1}=QV_{1}(\beta+QV_{1})$. The corresponding wave functions are given by
\begin{eqnarray}
\phi_{1}(s)=N_{1}s^{\alpha'/2}(1-s)^{\beta'/2}P_{n}^{(\alpha',\beta')}(1-2s)\,,
\end{eqnarray}
with $\alpha'=\frac{1}{\beta}\sqrt{m^{2}_{0}-E^{2}+2QV_{0}(E+m_{0})+Q^{2}V^{2}_{1}\,}$, and $\beta'=\frac{1}{\beta}\sqrt{m^{2}_{0}-E^{2}\,}$.

For the lower component ($i=2$), we have the following energy eigenvalue equation from (3.7) with the potential parameter $m_{2}=V_{0}-S_{0}$
\begin{eqnarray}
\left[\sqrt{m^{2}_{0}-E^{2}+2QV_{0}(E+m_{0})+a_{2}\,}+\sqrt{m^{2}_{0}-E^{2}\,}+\beta(2n+1)\right]^{2}\nonumber\\-\left[(-\beta+QV_{2})^{2}-4Q^{2}S_{0}(V_{0}-S_{0})\right]=0\,,
\end{eqnarray}
with $a_{2}=-4Qm_{0}S_{0}+Q^{2}V^{2}_{1}-4Q^{2}S_{0}(V_{0}-S_{0})$, and $b_{2}=-2Qm_{0}-4Q^{2}S_{0}(V_{0}-S_{0})+QV_{1}(\beta+QV_{1})$. The corresponding wave functions are written as
\begin{eqnarray}
\phi_{2}(s)=N_{2}s^{\alpha'/2}(1-s)^{\beta'/2}P_{n}^{(\alpha',\beta')}(1-2s)\,,
\end{eqnarray}
with $\alpha'=\frac{1}{\beta}\sqrt{m^{2}_{0}-E^{2}+2QV_{0}(E+m_{0})-4Qm_{0}S_{0}+Q^{2}V^{2}_{1}-4Q^{2}S_{0}(V_{0}-S_{0})\,}$, and $\beta'=\frac{1}{\beta}\sqrt{m^{2}_{0}-E^{2}\,}$. Before going further, we tend to give some numerical results obtained from Eqs. (3.15) and (3.17) in Table 1 where one observes that the energy values  for upper component larger than the ones for lower component, and numerical values for both components decrease while quantum number $n$ increase.

Now we can modify our results to the case where the mass is constant. Let us first write $m_{1}=0$ which means $V_{0}=-S_{0}$. This situation corresponds to the spin symmetric case for the Dirac equation in $3+1$ dimension [20-22]. We write the energy eigenvalue equation for the Dirac equation with the generalized Hulth\'{e}n potential as
\begin{eqnarray}
\left[\sqrt{m^{2}_{0}-E^{2}+2QV_{0}(E+m_{0})+a_{1}\,}+\sqrt{m^{2}_{0}-E^{2}\,}+\beta(2n+1)\right]^{2}-(\beta+QV_{1})^{2}=0\,,
\end{eqnarray}
with $a_{1}=Q^{2}V^{2}_{1}$, and $b_{1}=QV_{1}(\beta+QV_{1})$. Here, one has to choose the positive eigenvalues because in the case of the spin symmetry occurs only the bound states with positive energy [22]. For the constant mass, the wave functions with normalization constant given in Eq. (3.14) are
\begin{eqnarray}
\phi_{1}(s)=N_{1}s^{\alpha'/2}(1-s)^{\beta'/2}P_{n}^{(\alpha',\beta')}(1-2s)\,.
\end{eqnarray}
with $\alpha'=\frac{1}{\beta}\sqrt{m^{2}_{0}-E^{2}+2QV_{0}(E+m_{0})+Q^{2}V^{2}_{1}\,}$, and $\beta'=\frac{1}{\beta}\sqrt{m^{2}_{0}-E^{2}\,}$. The case where $m_{2}=0$ giving $V_{0}=+S_{0}$ corresponds to the pseudospin symmetric situation for the Dirac equation [20-22], and the energy eigenvalue equation becomes
\begin{eqnarray}
\left[\sqrt{m^{2}_{0}-E^{2}+2QV_{0}(E+m_{0})+a_{2}\,}+\sqrt{m^{2}_{0}-E^{2}\,}-\beta(2n+1)\right]^{2}-(\beta+QV_{1})^{2}=0\,,\nonumber\\
\end{eqnarray}
with $a_{2}=-4Qm_{0}S_{0}+Q^{2}V^{2}_{1}$, and $b_{2}=-4Qm_{0}S_{0}+QV_{1}(\beta+QV_{1})$. The last equation can give negative or positive eigenvalues, but one uses only negative energy eigenvalues because negative energy states can exist in the case of pseudospin symmetry [22]. The corresponding wave functions are given as
\begin{eqnarray}
\phi_{2}(s)=N_{2}s^{\alpha'/2}(1-s)^{\beta'/2}P_{n}^{(\alpha',\beta')}(1-2s)\,.
\end{eqnarray}
with $\alpha'=\frac{1}{\beta}\sqrt{m^{2}_{0}-E^{2}+2QV_{0}(E+m_{0})-4Qm_{0}S_{0}+Q^{2}V^{2}_{1}\,}$, and $\beta'=\frac{1}{\beta}\sqrt{m^{2}_{0}-E^{2}\,}$. The pseudospin symmetry, as a hidden symmetry in atomic nuclei, has been suggested firstly by Arima and co-workers [25, 26]. After the pseudospin symmetry has found a place as a relativistic symmetry in literature, some special features, spin symmetry for example, have been studied [27]. There have been many efforts about the recent progress on pseudospin and spin symmetry in different systems such as stable, exotic, deformed and spherical nuclei. These efforts extend the subject of "hidden symmetries" in atomic nuclei to include different perspectives such as perturbative study of the pseudospin symmetry, SUSY approach to hidden symmetries combining with similarity renormalization group and studying the source of some particular states which intrude from the major shell above to the shell below forming the nuclear magic numbers $28, 50, 82$, etc. [27].

Finally, we tend to give only the eigenvalue equation for the non-relativistic limit which can be obtained by using $E-m_{0} \sim E$ and $E+m_{0} \sim 2m_{0}$ in (3.7) ($\hbar=c=1$)
\begin{eqnarray}
&\left[\sqrt{-2m_{0}E+4Qm_{0}V_{0}+a_{i}\,}+\sqrt{-2m_{0}E\,}+\beta(2n+1)\right]^{2}\nonumber\\&-\left[(\beta+QV_{i})^{2}+Q^{2}[(m_{i}-S_{0})^{2}-V^{2}_{0}]\right]=0\,.\nonumber\\
\end{eqnarray}
The last equation gives two different results for energy eigenvalues, and one should choose the appropriate one.

\section{Bound States for $PT$-symmetric Forms}

Let us now study the case where the potential parameter $\beta$ is pure imaginary which means that the potential has a complex form as following
\begin{eqnarray}
V(x)=QV_{0}\,\frac{\cos(2\beta x)+i\sin(2\beta x)}{\cos(2\beta x)+i\sin(2\beta x)-Q}\,,
\end{eqnarray}
with $i=\sqrt{-1\,}$. This form of the potential in Eq. (1.1) is $PT$-symmetric because it satisfies
\begin{eqnarray}
\left[V(-x)\right]^{*}=V(x)\,.
\end{eqnarray}
which is non-Hermitian [4]. The bound state spectra of the generalized, $PT$-symmetric Hulth\'{e}n potential can be found from Eq. (3.7), and we write it explicitly as
\begin{eqnarray}
\sqrt{m^{2}_{0}-E^{2}+2QV_{0}(E+m_{0})+a_{i}\,}+\sqrt{m^{2}_{0}-E^{2}\,}+i\beta n'+\lambda\sqrt{\Gamma_{i}\,}=0\,,
\end{eqnarray}
with
\begin{eqnarray}
n'=2n+1\,\,; \Gamma_{i}=(i\beta+QV_{i})^{2}+Q^{2}[(m_{i}-S_{0})^{2}-V^{2}_{0}]\,.
\end{eqnarray}
where $\lambda=\pm 1$. The obtained result says that four different solution can be possible, and we expect that one of them, at least, gives a real spectra for the $PT$-symmetric Hulth\'{e}n potential [4]. The corresponding upper and lower components of the Dirac spinor are written with the help of Eq. (3.8) as
\begin{eqnarray}
\phi_{i}(s) \sim s^{\alpha''/2}(1-s)^{\beta''/2}P_{n}^{(\alpha'',\beta'')}(1-2s)\,.
\end{eqnarray}
where $\alpha''=-\alpha'$, and $\beta''=\frac{1}{\beta}\,\sqrt{E^{2}-m^{2}_{0}\,}=-\beta'$. We write the upper and lower spinor component without the normalization constant, but it can be computed in a similar way given in the above Section by using a modified normalization condition written for the non-Hermitian quantum systems [36-38].

An interesting form of the potential can be obtained if all potential parameters are taken pure imaginary, namely, $V_{0}\rightarrow iV_{0} (S_{0}\rightarrow iS_{0}), \beta \rightarrow i\beta, q \rightarrow iq$, giving
\begin{eqnarray}
V(x)=V_{0}\,\frac{q-\sin(\beta x)-i\cos(\beta x)}{q^2-2q\sin(\beta x)+1}=V^{*}\left(\frac{\pi}{2}-x\right)\,,
\end{eqnarray}
which is $PT$-symmetric but non-Hermitian (and also pseudo-Hermitian) [4, 36]. The energy spectra for this form of the potential is written as
\begin{eqnarray}
\sqrt{m^{2}_{0}-E^{2}+2QV_{0}(E+m_{0})+a_{i}\,}+\sqrt{m^{2}_{0}-E^{2}\,}+i\beta n'+\lambda\sqrt{\Gamma_{i}\,}=0\,,
\end{eqnarray}
with
\begin{eqnarray}
&n'=2n+1\,\,; \Gamma_{i}=-(\beta-QV_{i})^{2}+Q^{2}[(m_{i}-iS_{0})^{2}+V^{2}_{0}]\nonumber\\&a_{i}=-2m_{0}Q(V_{0}+S_{0})-2iQm_{0}m_{i}-Q^2[(m_{i}-iS_{0})^{2}+V^{2}_{0}]-Q^{2}V^{2}_{i}\,.
\end{eqnarray}
It is worthwhile to say that one has to chose the result giving a real spectrum obtained from Eq. (4.7) for the above form of the generalized Hulth\'{e}n potential.

\section{Solutions for Specific $q$-values}

The value of $q=+1$ corresponds to the standard Hulth\'{e}n potential for which the energy equation is obtained from Eq. (3.7), and the upper and lower components of Dirac spinor from Eq. (3.8). For $q=-1$, the generalized Hulth\'{e}n potential gives
\begin{eqnarray}
V(x)=-V_{0}\,\frac{e^{-2\beta x}}{e^{-2\beta x}+1}\,,
\end{eqnarray}
which is the Woods-Saxon potential. The energy levels and upper and lower components of Dirac spinor for this form are obtained from Eqs. (3.7) and (3.8), respectively.

For $q=0$, we have
\begin{eqnarray}
V(x)=-V_{0}e^{-2\beta }\,,
\end{eqnarray}
which is the exponential potential, and it is known that there is no explicit expression for the bound states for non-relativistic, and relativistic wave equations [39-41]. Hence we have to reconsider the problem by using the new variable $s=e^{-2\beta x}$ giving
\begin{eqnarray}
\left\{\frac{d^2}{ds^2}+\frac{1}{s}\frac{d}{ds}-\frac{1}{s^2}\left[\frac{A'}{4\beta^2}-\frac{2B'}{4\beta^2}s+\frac{C'}{4\beta^2}s^2\right]\right\}\phi_{i}(s)=0\,,
\end{eqnarray}
with
\begin{eqnarray}
&&A'=m^{2}_{0}-E^{2}\,,\nonumber\\
&&B'=m_{0}m_{i}-m_{0}S_{0}+EV_{0}+\beta V_{i}\,,\nonumber\\
&&C'=V^{2}_{i}+(m_{i}-S_{0})^{2}-V^{2}_{0}\,,
\end{eqnarray}
We compare Eq. (5.3) with Eq. (A.1) in Appendix, and with the help of Eq. (A.3) we obtain the parameter set
\begin{eqnarray}
\alpha_{1}=1, \alpha_{2}=\alpha_{3}=0,\xi_{1}=\frac{C'}{4\beta^2},\xi_{2}=\frac{2B'}{4\beta^2},\xi_{3}=\frac{A'}{4\beta^2},\nonumber\\
\alpha_{4}=\alpha_{5}=0,\alpha_{6}=\xi_{1},\alpha_{7}=-\xi_{2},\alpha_{8}=\xi_{3},\alpha_{9}=\xi_{1},\nonumber\\
\alpha_{10}=1+2\sqrt{\xi_{3}\,},\alpha_{11}=2\sqrt{\xi_{1}\,},
\alpha_{12}=\sqrt{\xi_{3}\,},\alpha_{13}=\sqrt{\xi_{1}\,}\,,
\end{eqnarray}

Eq. (A.10) gives the upper and lower component of Dirac spinor for exponential potential in Eq. (5.2)
\begin{eqnarray}
\phi_{i}(s) \sim s^{\frac{1}{2\beta}\sqrt{m^{2}_{0}-E^{2}\,}}e^{\frac{1}{2\beta}\sqrt{(m_{i}-S_{0})^{2}-V^{2}_{0}+V^{2}_{i}\,}}\,L_{n}^{\frac{1}{\beta}\sqrt{m^{2}_{0}-E^{2}\,}}
(2\sqrt{\xi_{1}\,}s)\,.
\end{eqnarray}

\section{Conclusions}

We have analyzed the analytical solutions of the Dirac equation with scalar-vector-pseudoscalar generalized Hulth\'{e}n potential in $1+1$ dimension within the position-dependent mass formalism. We have reduced the two extended effective-mass versions of coupled equations written for the upper and lower component to a form of analytical solvable equations by relating the mass function with the potentials. We have given both energy eigenvalue equations and normalized wave functions in closed forms. We have also computed the results for the case where the mass is constant which correspond to spin and pseudospin symmetric cases in Dirac equation. We have written the results for the bound states in the non-relativistic case. We have studied the bound state spectrum and the corresponding normalized upper and lower component of Dirac spinor for the complex, generalized Hulth\'{e}n potential which are $PT$-symmetric, non-Hermitian forms of the potential.

\textbf{Competing Interests}

The author(s) declare(s) that there is no conflict of interests
regarding the publication of this paper.

\section{Acknowledgments}
The author thanks Prof Dr Andreas Fring from City University London and the Department of Mathematics for hospitality. This research was partially supported through a fund provided by University of Hacettepe.

\begin{appendices}
\section{}

The general form of a second order differential equation which is solved by using the parametric generalization of the Nikiforov-Uvarov method [23]
\begin{eqnarray}
\frac{d^{2}F(s)}{ds^{2}}+\frac{\alpha_{1}-\alpha_{2}s}{s(1-\alpha_{3}s)}\frac{dF(s)}{ds}-\frac{\xi_{1}s^{2}-\xi_{2}s+\xi_{3}}{[s(1-\alpha_{3}s)]^{2}}\,F(s)=0\,,
\end{eqnarray}
with the quantization rule
\begin{eqnarray}
\alpha_{2}n-(2n+1)\alpha_{5}+(2n+1)(\sqrt{\alpha_{9}\,}+\alpha_{3}\sqrt{\alpha_{8}\,})+n(n-1)\alpha_{3}+\alpha_{7}+2\alpha_{3}\alpha_{8}+2\sqrt{\alpha_{8}\alpha_{9}\,}=0\,,\nonumber\\
\end{eqnarray}
where $n=0, 1, 2, \ldots$.

The parameters $\alpha_{i}'s$ within this approach are defined as
\begin{eqnarray}
&&\alpha_{4}=\frac{1}{2}(1-\alpha_{1});\,\,\alpha_{5}=\frac{1}{2}(\alpha_{2}-2\alpha_{3});\,\,\alpha_{6}=\alpha^{2}_{5}+\xi_{1};\,\,\alpha_{7}=2\alpha_{4}\alpha_{5}-\xi_{2};\nonumber\\&&\alpha_{8}=\alpha^{2}_{4}+\xi_{3};\,\,\alpha_{9}=\alpha_{3}(\alpha_{7}+\alpha_{3}\alpha_{8})+\alpha_{6}\,,
\end{eqnarray}

The corresponding wave functions are given in terms of the parameters  $\alpha_{i}$ [23]
\begin{eqnarray}
F(s)=N\,s^{\alpha_{12}}(1-\alpha_{3}s)^{-\alpha_{12}-\frac{\alpha_{13}}{\alpha_{3}}}\,P_{n}^{(\alpha_{10}-1,\,\frac{\alpha_{11}}{\alpha_{3}}-\alpha_{10}-1)}(1-2\alpha_{3}s)\,,
\end{eqnarray}
where
\begin{eqnarray}
&&\alpha_{10}=\alpha_{1}+2\alpha_{4}+2\sqrt{\alpha_{8}\,};\,\,\alpha_{11}=\alpha_{2}-2\alpha_{5}+2(\sqrt{\alpha_{9}\,}+\alpha_{3}\sqrt{\alpha_{8}\,});\nonumber\\&&\alpha_{12}=\alpha_{4}+\sqrt{\alpha_{8}\,};\,\,\alpha_{13}=\alpha_{5}-(\sqrt{\alpha_{9}\,}+\alpha_{3}\sqrt{\alpha_{8}\,})\,.
\end{eqnarray}
with the Jacobi polynomials $P_{n}^{(\sigma_{1}, \sigma_{2})}(s)$, and a normalization constant $N$.

For the second independent solution the quantization condition is given by
\begin{eqnarray}
\alpha_{2}n+(1-2n)\alpha_{5}+(2n+1)(\sqrt{\alpha_{9}\,}-\alpha_{3}\sqrt{\alpha_{8}\,})+n(n-1)\alpha_{3}+\alpha_{7}+2\alpha_{3}\alpha_{8}-2\sqrt{\alpha_{8}\alpha_{9}\,}=0\,,\nonumber\\
\end{eqnarray}
with the corresponding wave functions
\begin{eqnarray}
F(s)=N\,s^{\alpha^{*}_{12}}(1-\alpha_{3}s)^{-\alpha^{*}_{12}-\frac{\alpha^{*}_{13}}{\alpha_{3}}}\,P_{n}^{(\alpha^{*}_{10}-1,\,\frac{\alpha^{*}_{11}}{\alpha_{3}}-\alpha_{10}-1)}(1-2\alpha_{3}s)\,,
\end{eqnarray}
where
\begin{eqnarray}
&&\alpha^{*}_{10}=\alpha_{1}+2\alpha_{4}-2\sqrt{\alpha_{8}\,};\,\,\alpha^{*}_{11}=\alpha_{2}-2\alpha_{5}-2(\sqrt{\alpha_{9}\,}-\alpha_{3}\sqrt{\alpha_{8}\,})\,,\nonumber\\&&\alpha^{*}_{12}=\alpha_{4}-\sqrt{\alpha_{8}\,};\,\,\alpha^{*}_{13}=\alpha_{5}-(\sqrt{\alpha_{9}\,}-\alpha_{3}\sqrt{\alpha_{8}\,})\,.
\end{eqnarray}

If a situation appearing in the problem such as $\alpha_{3}=0$, then the quantization rule in (A.2) becomes
\begin{eqnarray}
(\alpha_{2}-2\alpha_{5})n+(2n+1)(\sqrt{\alpha_{9}\,}-\alpha_{3}\sqrt{\alpha_{8}\,})&+&n(n-1)
\alpha_{3}+\alpha_{7}\nonumber\\&+&2\alpha_{3}\alpha_{8}-2\sqrt{\alpha_{8}\alpha_{9}\,}+\alpha_{5}=0\,,
\end{eqnarray}
with the corresponding wave functions
\begin{eqnarray}
F(s)=Ns^{\alpha_{12}}e^{\alpha_{13}\,s}L_{n}^{\alpha_{10}-1}(\alpha_{11}\,s)\,.
\end{eqnarray}
when the limits become $\lim_{\alpha_{3}\rightarrow 0}\,P_{n}^{(\alpha_{10}-1,\,\frac{\alpha_{11}}{\alpha_{3}}-\alpha_{10}-1)}(1-2\alpha_{3}s)=L_{n}^{\alpha_{10}-1}(\alpha_{11}\,s)$ and $\lim_{\alpha_{3}\rightarrow 0}(1-\alpha_{3}s)^{-\alpha_{12}-\frac{\alpha_{13}}{\alpha_{3}}}=e^{\alpha_{13}\,s}$ with generalized Laguerre polynomials $L_{n}^{\sigma_{3}}(s)$.
\end{appendices}

\newpage

\begin{table}
\begin{center}
\begin{tabular}{ccc}
\hline
$n$ & energy values for $i=1$ & energy values for $i=2$ \\
\hline
0 & -15.97700 & -28.37410 \\
\hline
1 & -17.93500 & -30.26090 \\
\hline
2 & -19.78040 & -32.01920 \\
\hline
3 & -21.52680 & -33.66200 \\
\hline
4 & -23.18510 & -35.19950 \\
\hline
\end{tabular}
\end{center}
\caption{The variation of energy eigenvalues with different $n$ for $\beta=1, m_0=50, V_1=V_2=1.5, Q=100, V_0=1, S_0=2$.}
\end{table}

\end{document}